\documentstyle[12pt]{article}
\begin{document}
\begin{center}{\large\bf ON THE EFFECT OF BINDING IN DEEP
INELASTIC SCATTERING}\end{center}
\vskip 1em \begin{center} {\large Felix M. Lev} \end{center}
\vskip 1em \begin{center} {\it Laboratory of Nuclear
Problems, Joint Institute for Nuclear Research, Dubna, Moscow region
141980 Russia (E-mail:  lev@nusun.jinr.dubna.su)} \end{center}
\vskip 1em
\begin{abstract}
 The phenomenon of scaling in deep inelastic lepton-nucleon scattering
is usually explained in terms of the Feynman parton model, and the
logarithmic corrections to scaling are explained in the framework of
perturbative QCD. For testing the validity of the parton model,
we consider the deep inelastic electron scattering in a model in which
the system electromagnetic current operator explicitly satisfies
relativistic invariance
and current conservation. Let the struck particle have the fraction $\xi$
of the total momentum in the infinite momentum frame. Then it is shown
that, due to binding of particles in the system under consideration,
the Bjorken variable $x$ no longer can be interpreted as $\xi$, even in
the Bjorken limit and in zero order of the perturbation theory. We argue
that, as a result, the data on deep inelastic scattering alone do not
make it possible to determine the $\xi$ distribution of quarks in the
nucleon.
\end{abstract}
\vskip 1em

\section{Introduction}
\label{S1}

 The phenomenon of scaling in deep inelastic scattering (DIS) was first
explained in the framework of approach developed by Bjorken \cite{Bjor}.
Another explanation was proposed by Feynman \cite{Feyn} in the framework
of the parton model. According to this model, the process of absorption
of a virtual photon with the 4-momentum $q$ such that $|q^2|$ is very
large, can be described assuming that the nucleon consists of point-like
partons which do not interact with each other at small
distances. Let $P'$ be the 4-momentum of the nucleon in the infinite
momentum frame (IMF) where the nucleon moves along the positive direction
of the $z$ axis with the velocity close to the velocity of light. Let also
the struck parton have the fraction $\xi$ of the nucleon's momentum.
Then the 4-momentum of this parton in the
final state is equal to $\xi P'+q$. We consider the process in the Bjorken
limit when $|q^2|$ and $2(P'q)$ are very large, but the quantity
$x=|q^2|/2(P'q)$ is not too close to 0 or 1. Then assuming that
$(\xi P'+q)^2$ does not
exceed the square of the nucleon's mass, we conclude that in the Bjorken
limit $\xi = x$. For this reason the authors of some textbooks and papers
even do not distinguish the quantities $\xi$ and $x$.

 In QCD the partons are naturally identified with quarks, and the fact
that they do not interact at small distances is treated as the consequence
of the asymptotic freedom. More exactly, using some assumptions, it can
be shown that in the Bjorken limit $\xi=x$ and Bjorken Scaling take
place in zero order in $\alpha_s$ (where $\alpha_s$ is the QCD running
coupling constant), while the interaction between quarks and gluons
can be taken into account perturbatively (leading to logarithmic
breaking of scaling and the relation $\xi=x$). Note however that the
technique of the operator product expansion (OPE)
developed by Wilson and others (see, for example, Refs. \cite{Wil})
formally does not use any relation between $\xi$ and $x$, and in principle
one cannot exclude the possibility that, even in the Bjorken limit and in
zero order of the perturbation theory, scaling takes place while
$\xi \neq x$. The relation $\xi=x$ in the framework of the OPE can be
obtained only at some additional assumptions---see Refs. \cite{mon}.
Moreover, it is not even clear whether the OPE series is convergent or
asymptotic \cite{Greenberg}.

 The problem arises how to take into account the effect that initially
quarks are in the bound state --- in the nucleon. It is clear that
this effect cannot be considered in the framework of perturbative QCD.

\begin{sloppypar}
 Our experience in nonrelativistic quantum mechanics and in nuclear
physics tells that the effect of binding is not important at large
momentum transfer. For example, the results of calculations of the
electron scattering from nuclei show that at small momentum transfer
there is a coherent process on the nucleus as a whole (and therefore
the scattering amplitude is proportional to $Z$ and the cross-section
is proportional to $Z^2$), while at large momentum transfer the
cross-section is an incoherent sum of the cross-sections on each
nucleon in the nucleus (therefore, this cross-section is proportional
to $Z$).
\end{sloppypar}

 This picture was questioned by many authors after the discovery of the
EMC effect \cite{EMC}. The central point of the extensive discussion
in the literature was whether the EMC effect can be explained in the
framework of conventional nuclear physics. We shall not discuss
this problem but note that the large group of authors stated that this
can be done if the effects of the interaction between the nucleons and
relativistic effects are taken into account (see, for
example the calculations in Refs. \cite{Ak,FS,Ciofi}).

It is important to note that the above calculations in conventional
nuclear physics (not taking into account the EMC effect) and the
analogous calculations in atomic physics have been carried out in the
framework of the impulse approximation, where it is assumed that the
electromagnetic current operator (ECO) of the system under consideration
can be represented as a sum of the ECO's for the constituents comprising
this system. Such an approximation is reasonable in nonrelativistic
quantum mechanics, but it is well-known that in the relativistic case
the ECO should necessarily contain the terms depending on the interaction
between the constituents, since otherwise the ECO satisfies neither
relativistic invariance nor current conservation.

 In the parton model it is assumed that the partons in the IMF are free
to the extent that the impulse approximation is valid. On the other hand,
the interaction between them cannot be eliminated at all since in this
case the nucleon will not be bound. Are these assumptions compatible
with each other? The answer to this question can be given only in the
framework of explicitly solvable models.

 In the present paper the effect of binding in DIS is investigated in
the framework of the model in which the ECO explicitly satisfies
relativistic invariance, current conservation,
cluster separability, and the condition that the interaction terms
in the ECO do not renormalize the total system electric charge. However
these conditions are not sufficient for choosing a unique solution. We
choose a special solution considered in Ref. \cite{lev}. The essence of
our results becomes clear already in the case of $N=2$ particles, and
then these results are generalized to the case when $N$ is arbitrary
(including $N=\infty$).

 The major objection against such an approach maybe that the
ECO ${\hat  J}^{\mu}(x)$ ($\mu =0,1,2,3$) obtained in such a way is
nonlocal in the sense that it is not derived from a local Lagrangian
(we use $x$ to denote a point in Minkowski space as well as the Bjorken
variable, but this should not lead to misunderstanding).
In particular, it is not clear whether the commutator
$[{\hat J}^{\mu}(x),{\hat J}^{\mu}(0)]$ necessarily vanishes when $x$
is a space-like vector.
Let us note however that if a theory is nonlocal in the above sense,
this does not necessarily imply that it is unphysical. Indeed, as it
has become clear already in 30th, in relativistic quantum theory there
is no operator possessing all the properties of the position operator.
In particular, the quantity $x$ in the Lagrangian density $L(x)$ is
not the coordinate, but some parameter which becomes the coordinate only
in the classical limit. Therefore the physical condition is that the
above commutator should vanish when $|x^2|\rightarrow \infty$ but $x^2<0$.
We shall see in Sec. \ref{S4} that this condition is indeed satisfied.

 Anyway, the relation between $\xi$ and $x$ derived in Sec. \ref{S5} is
in fact kinematical. This relation shows that $\xi \neq x$ even in the
Bjorken limit and in zero order of the perturbation theory.
At the same time, in our approach the scaling and the Callan-Gross
relation \cite{CallGr} remain.

 The paper is organized as follows. In Sec. \ref{S2} we give a detailed
calculation of the hadronic tensor for systems of two particles in the
impulse approximation, and in Sec. \ref{S3} the same is done for systems
of $N$ particles. The corresponding results are well-known, and the
analogous calculations were carried out elsewhere (see, for example,
Refs. \cite{Close,ILH,Web} and references cited therein), but the
major purpose
of these sections is to prepare the reader for the consideration of the
case when the ECO contains the interaction. In Sec. \ref{S4} we briefly
describe the results of Ref. \cite{lev} needed in Sec. \ref{S5} where
these results are used for the explicit calculation of the hadronic
tensor for systems of $N$ particles. We hope that the main part of the
paper is self-contained, and even the unexperienced reader can follow
our calculations.

\section{Impulse approximation for the system of two particles}
\label{S2}

Let us consider a system of two particles with the masses $m_i$ and
the electric charges $e_i$ ($i=1,2$). If $p$ is the 4-momentum of
some particle then ${\bf p}_{\bot}$ means the projection of $p$ onto
the plane $xy$, and, instead of the temporal and the $z$ components
of $p$, we use the $\pm$ components defined as $p^{\pm}=
(p^0\pm p^z)/\sqrt{2}$. We also use $\sigma_i$ to denote the
projection of the spin of particle $i$ on the $z$ axis.

 The Hilbert space $H$ for the system under consideration is the space
of functions $\varphi({\bf p}_{1\bot},p_1^+,\sigma_1,
{\bf p}_{2\bot},p_2^+,\sigma_2)$ such that
\begin{equation}
\sum_{\sigma_1\sigma_2}\int\nolimits ||\varphi({\bf p}_{1\bot},p_1^+,
\sigma_1,{\bf p}_{2\bot},p_2^+,\sigma_2)||^2
d\rho({\bf p}_{1\bot},p_1^+)d\rho({\bf p}_{2\bot},p_2^+)\quad< \infty
\label{1}
\end{equation}
where
\begin{equation}
d\rho({\bf p}_{\bot},p^+)= \frac{d^2{\bf p}_{\bot}dp^+}
{2(2\pi)^3p^+}
\label{2}
\end{equation}

 Instead of the individual particle variables we introduce the total
momentum variables and the internal momentum variables. The former
are the $\bot$ and + components of the 4-vector $P=p_1+p_2$, and,
following Ref. \cite{Ter}, the latter can be defined as
\begin{equation}
\xi=\frac{p_1^+}{P^+},\quad {\bf k}_{\bot}={\bf p}_{1\bot}-\xi P^+
\label{3}
\end{equation}
As shown in this reference, it is also possible to choose as the
internal variables the set ${\bf k}=({\bf k}_{\bot},k^z)$ where $k^z$
is defined from the conditions
\begin{equation}
\xi=\frac{\omega_1({\bf k})+k^z}{M({\bf k})},
\label{4}
\end{equation}
$\omega_i({\bf k})=(m_i^2+{\bf k}^2)^{1/2}$, and $M({\bf k})=
\omega_1({\bf k})+\omega_2({\bf k})$. It is easy to show that $P^2=
M({\bf k})^2$, and therefore the mass operator of the two-particle
system is the operator of multiplication by $M({\bf k})$.

 It is easy to see that under the interchange of particles 1 and 2,
${\bf k}\rightarrow -{\bf k}$ and $\xi \rightarrow 1-\xi$. For this
reason it is sometimes convenient to use the notations ${\bf k}_1=
-{\bf k}_2={\bf k}$ and $\xi_1=1-\xi_2=\xi$.

 Let us introduce the 4-vectors $k_i=(\omega_i({\bf k}_i),{\bf k}_i)$
and $G=P/M({\bf k})$. Since $G^2=1$, only three components of $G$ are
independent, for example ${\bf G}_{\bot}$ and $G^+$. Let $\beta(G)
\equiv \beta({\bf G}_{\bot},G^+)\in SL(2,C)$ be the matrix with the
components
\begin{equation}
\beta_{11}=\beta_{22}^{-1}=2^{1/4}(G^+)^{1/2},\quad \beta_{12}=0,\quad
\beta_{21}=(G^x+\imath G^y)\beta_{22}
\label{5}
\end{equation}

 We use $L(l)$ to denote the Lorentz transformation corresponding
to $l\in SL(2,C)$. Then a direct calculation shows that
\begin{equation}
p_i=L[\beta(G)]k_i \quad (i=1,2)
\label{6}
\end{equation}
Therefore $L[\beta(G)]$ has the meaning of the boost, and ${\bf k}$ is
the momentum in the c.m. frame. We note that these quantities are the
same as the "canonical" ones if ${\bf G}_{\bot}=0$.

 A direct calculation gives
\begin{eqnarray}
&&d\rho({\bf p}_{1\bot},p_1^+)d\rho({\bf p}_{2\bot},p_2^+)=
d\rho({\bf P}_{\bot},P^+)d\rho(int),\nonumber\\
&&d\rho(int)=\frac{d^2{\bf k}_{\bot}d\xi}{2(2\pi)^3\xi(1-\xi)}
=\frac{M({\bf k})d^3{\bf k}}{2(2\pi)^3\omega_1({\bf k})
\omega_2({\bf k})}
\label{7}
\end{eqnarray}

\begin{sloppypar}
 We introduce the Hilbert space $H_{int}$ as the space of functions
$\chi({\bf k}_{\bot},\xi,\sigma_1,\sigma_2)\equiv
\chi({\bf k},\sigma_1,\sigma_2)$ such that
\begin{equation}
||\chi||^2\equiv \sum_{\sigma_1\sigma_2}\int\nolimits
|\chi({\bf k}_{\bot},\xi,\sigma_1,\sigma_2)|^2d\rho(int)\, < \, \infty
\label{8}
\end{equation}
(note that $\xi \in [0,1]$). We shall write the function $\chi$ in the
form $\chi({\bf k}_{2\bot},\xi_2,\sigma_2,\sigma_1)$ if
$({\bf k}_{2\bot},\xi_2)$ are chosen as the independent variables, and
$({\bf k}_{1\bot},\xi_1)$ are connected with them as explained above.
\end{sloppypar}

 In the scattering theory the state in which particle 1 has the momentum
$p_1'$ and the spin projection $\sigma_1'$, and particle 2 has the
momentum $p_2'$ and the spin projection $\sigma_2'$, is the product
$|p_1',\sigma_1'\rangle |p_2',\sigma_2'\rangle$ where
\begin{equation}
|p_i',\sigma_i'\rangle =2(2\pi)^3p_i^{'+}\delta^{(2)}({\bf p}_{i\bot}-
{\bf p}_{i\bot}')\delta(p_i^+-p_i^{'+})\delta_{\sigma_i\sigma_i'}
\label{9}
\end{equation}
($\delta_{\sigma_i\sigma_i'}$ is the Cronecker symbol). If the particles
are in the bound state described by the wave function $\chi' \in H_{int}$,
and the system as the whole has the 4-momentum $P'$ then, as shown by
several authors (see, for example, Ref. \cite{lev1}), the above choice
of the variables makes it possible to write the wave function of such a
system  by analogy with Eq. (\ref{9}):
\begin{equation}
|P',\chi'\rangle = 2(2\pi)^3P^{'+}
\delta^{(2)}({\bf P}_{\bot}-{\bf P}_{\bot}')\delta(P^+-P^{'+})\chi'
\label{10}
\end{equation}
where $\chi'$ is normalized as $||\chi'||^2=1$.

 We shall always assume that all particles having the electric charge
are structureless and their spin is equal to 1/2. Then the one-particle
ECO for particle $i$ acts over the variables of this particle as
\begin{eqnarray}
J_i^{\mu}\varphi({\bf p}_{i\bot},p_i^+,\sigma_i)&=&
r_i\sum_{\sigma_i'}\int\nolimits [\bar{w}_i(p_i,\sigma_i)\gamma^{\mu}
w_i(p_i',\sigma_i')]\cdot\nonumber\\
&&\varphi({\bf p}_{i\bot}',p_i^{'+},
\sigma_i')d\rho({\bf p}_{i\bot}',p_i^{'+})
\label{11}
\end{eqnarray}
and over the variables of other particle it acts as the identity
operator. Here $r_i=e_i/e_0$ is the ratio of the particle electric
charge to the unit electric charge, $w_i(p_i,\sigma_i)$ is the Dirac light
cone spinor, $\gamma^{\mu}$ is the
Dirac $\gamma$-matrix, and $\bar{w}=w^{+}\gamma^0$. The form of
$w_i(p_i,\sigma_i)$ in the spinor representation of the Dirac
$\gamma$-matrices is
\begin{equation}
w_i(p_i,\sigma_i)=\sqrt{m_i}
\left\|\begin{array}{c}
\beta({\bf p}_{i\bot}/m_i,p_i^+/m_i)\chi(\sigma)\\
\beta({\bf p}_{i\bot}/m_i,p_i^+/m_i)^{-1+}\chi(\sigma)
\end{array}\right\|
\label{12}
\end{equation}
where $\chi(\sigma)$ is the ordinary spinor describing the state with
the spin projection on the $z$ axis equal to $\sigma$.

 The impulse approximation implies that the ECO for the system as a
whole is a sum for of the ECO's for the constituents. In particular, for
a system of two particles $J^{\mu}(0)=J_1^{\mu}(0)+J_2^{\mu}(0)$. This
relation together with Eqs. (\ref{9}-\ref{11}) makes it possible to
calculate the matrix element of the operator $J^{\mu}(0)$ between any
two-particle states.

 When the bound state of two particles absorbs a virtual photon with
the large momentum, we should expect that, since the relative momentum
in the final state is large, the interaction between the particles in
this state can be neglected. Therefore the inclusive cross-section is
fully defined by the tensor
\begin{eqnarray}
&&W^{\mu\nu}=\frac{1}{4\pi}\sum_{\sigma_1"\sigma_2"}\int\nolimits
(2\pi)^4\delta^{(4)}(P'+q-p_1"-p_2")
\langle P',\chi'|J^{\mu}(0)|p_1",\sigma_1", \nonumber\\
&&p_2",\sigma_2"\rangle
\langle p_1",\sigma_1",p_2",\sigma_2"|J^{\nu}(0)|P',\chi'\rangle
d\rho({\bf p}_{1\bot}",p_1^{"+})d\rho({\bf p}_{2\bot}",p_2^{"+})
\label{13}
\end{eqnarray}

 It is well-known that the average value of this tensor over the all
initial spin states is equal to
\begin{eqnarray}
&&W^{\mu\nu}(P',q)=(\frac{q^{\mu}q^{\nu}}{q^2}-g^{\mu\nu})
F_1(x,q^2)+\nonumber\\
&&\frac{1}{(P'q)}(P^{'\mu}-\frac{q^{\mu}(P'q)}{q^2})
(P^{'\nu}-\frac{q^{\nu}(P'q)}{q^2})F_2(x,q^2),
\label{14}
\end{eqnarray}
and our goal is to calculate the structure functions $F_1(x,q^2)$ and
$F_2(x,q^2)$.

 Knowing the momenta $p_i"$ in the final state and using Eq. (\ref{6})
we can calculate the relative momentum ${\bf k}"$ in the final state.
Let $M"=\omega_1({\bf k}")+\omega_2({\bf k}")$ be the mass of the
final state. Then a standard calculation gives
\begin{equation}
(2\pi)^4\delta^{(4)}(P'+q-p_1"-p_2")d\rho({\bf p}_{1\bot}",p_1^{"+})
d\rho({\bf p}_{2\bot}",p_2^{"+})=\frac{k"do"}{16\pi^2M"}
\label{15}
\end{equation}
where $k"=|{\bf k}"|$, and $do"$ is the element of the solid angle for
the unit vector ${\bf k}"/k"$. In the Bjorken limit $k"=M"/2$ and
\begin{equation}
M"=[\frac{|q^2|(1-x)}{x}]^{1/2}
\label{16}
\end{equation}
since $M^{"2}=(P'+q)^2$.

 As follows from Eqs. (\ref{6}), (\ref{7}), (\ref{9}) and (\ref{11})
\begin{eqnarray}
&&\langle p_1^",\sigma_1^",p_2^",\sigma_2^"|J^{\nu}(0)|
P',\chi'\rangle=\nonumber\\
&&\sum_{i=1}^{2}\sum_{\sigma_i} \frac{r_i}{\xi_i}
[\bar{w}_i(p_i",\sigma_i")\gamma^{\nu}w_i(p_i',\sigma_i')]
\chi'({\bf d}_i,\sigma_i',\tilde{\sigma}_i")
\label{17}
\end{eqnarray}
where $\tilde{\sigma}_1"=\sigma_2"$, $\tilde{\sigma}_2"=\sigma_1"$,
$p_i'=L[\beta({\bf P}_{\bot}'/M({\bf d}_i),P^{'+}/M({\bf d}_i))]d_i$,
$d_i=(\omega_i({\bf d}_i),{\bf d}_i)$, and the vectors ${\bf d}_i$ are
defined by the conditions
\begin{equation}
L[\beta(\frac{{\bf P}_{\bot}'}{M({\bf d}_i)},
\frac{P^{'+}}{M({\bf d}_i)})](\tilde{\omega}_i({\bf d}_i),-{\bf d}_i)=
L[\beta(\frac{{\bf P}_{\bot}"}{M"},\frac{P^{"+}}{M"})]
(\tilde{\omega}_i({\bf k}_i"),-{\bf k}_i")
\label{18}
\end{equation}
where $P"=p_1"+p_2"$, $\tilde{\omega}_1=\omega_2$, and
$\tilde{\omega}_2=\omega_1$.

 It is convenient to consider the process in the reference frame where
${\bf P}_{\bot}'={\bf q}_{\bot}=0$, and $P^{'z}$ is positive and very
large. By analogy with the Breit frame for elastic processes we choose
the reference frame in which ${\bf P}'+{\bf P}"=0$. It is easy to show
that in this reference frame
\begin{equation}
q^0=2|{\bf P}'|(1-x),\, P^{'+}=\sqrt{2}|{\bf P}'|,\,
q^+=-\sqrt{2}|{\bf P}'|x, \, P^{"+}=\sqrt{2}|{\bf P}'|(1-x)
\label{19}
\end{equation}
Then, as follows from Eq. (\ref{18})
\begin{eqnarray}
{\bf d}_{1\bot}={\bf k}_{1\bot}",\,&& {\bf d}_{2\bot}={\bf k}_{2\bot}",\quad
1-\xi_1=\frac{1}{2}(1-x)(1-cos\theta),\nonumber\\
&&1-\xi_2=\frac{1}{2}(1-x)(1+cos\theta)
\label{20}
\end{eqnarray}
where $k^{"z}=k"cos\theta$, and the quantities $\xi_i$ are expressed
in terms of ${\bf d}_i$ according to Eq. (\ref{4}).

 We assume that the internal wave function $\chi'({\bf d})$ effectively
cuts the contribution of large momenta, and therefore the contribution
to the integrals containing $\chi'({\bf d})$ is given only by the
momenta with $|{\bf d}|\leq m_0$, where $m_0$ is some parameter satisfying
the condition $m_0^2 \ll |q^2|$. Then, as follows from Eqs.
(\ref{15}-\ref{17}) and the first pair of expressions in Eq. (\ref{20}),
only those $\theta's$ contribute to Eq. (\ref{13}) for which
$1-|cos\theta|\leq m_0^2/|q^2|$. Using additionally Eq. (\ref{4}), the
second pair of expressions in Eq. (\ref{20}), and the condition
$|d^z|\leq m_0$, we conclude that in the right-hand side of Eq. (\ref{17})
the term with $i=1$ is not negligible only if $cos\theta$ is close to -1
($1+cos\theta\leq m_0^2/|q^2|)$, while the term with $i=2$ is not
negligible only if $cos\theta$ is close to +1
($1-cos\theta\leq m_0^2/|q^2|)$.

 Therefore, as follows from Eq. (\ref{20}), $\xi_1=x$ in the first term,
and $\xi_2=x$ in the second one, in agreement with the interpretation of
the quantity $x$ in the parton model. We also see that both particles
absorb the virtual photon incoherently.

 It is easy to see that in both regions of $cos\theta$ we can write
$do"=d^2{\bf k}_{\bot}"/k^{"2}$. Therefore, as follows from Eqs.
(\ref{5}-\ref{7}), (\ref{12}), (\ref{13}), (\ref{15}-{17}), (\ref{19})
and (\ref{20})
\begin{equation}
W^{\mu\nu}=\sum_{i=1}^{2} r_i^2 \int\nolimits \langle
\chi'({\bf k}_{i\bot},\xi_i=x)|S_i^{\mu\nu}|
\chi'({\bf k}_{i\bot},\xi_i=x)\rangle
\frac{d^2{\bf k}_{i\bot}}{4(2\pi)^3x(1-x)}
\label{21}
\end{equation}
where we do not write the spin variables in the arguments of the
function $\chi'$, the scalar product is taken over these variables, and
the tensor operator $S_i^{\mu\nu}$ is as follows. It is equal to zero
if either $\mu$ or $\nu$ is equal to $\pm$, while if $j,l=x,y$ then
$S_i^{jl}=\delta_{jl}+2\imath \epsilon_{jl} s_i^z$, where
$\epsilon_{jl}$ is the antisymmetric tensor with
$\epsilon_{xy}=-\epsilon_{yx}=1$, $\epsilon_{11}=\epsilon_{22}=0$, and
$s_i^z$ is the $z$ component of the spin operator for particle $i$.

\section{Impulse approximation for the system of $N$ particles}
\label{S3}

 If the bound state consists of $N$ particles, we can choose any pair
of particles, say the pair of particles $i_1$ and $i_2$, and construct
the external and internal variables for this pair as described above.
Let $P_{i_1i_2}=p_{i_1}+p_{i_2}$ be the total 4-momentum of the pair
$i_1i_2$. Using the analogous procedure, we can construct from
$P_{i_1i_2}$ and $p_{i_3}$ the total momentum of the system $i_1i_2i_3$
and the relative variables describing the motion of the system $i_1i_2$
relative particle $i_3$. Then the internal variables in the system
$i_1i_2i_3$ are these variables and the internal variables for the
system $i_1i_2$. It is obvious that in such a way we can construct the
external and internal variables for a system with any number of particles,
but the choice of the internal variables is not unique.

 Let $P_i$ be the total 4-momentum of the system consisting of
particles $1,2,...i-1,i+1,...N$, $\tilde{int}$ be a set of the internal
variables for this system, and $d\rho(\tilde{int})$ be the volume element
in the set $\tilde{int}$. Then $P=p_i+P_i$ is the
total 4-momentum of the system consisting of all the particles $1,2,...N$,
and we use ${\bf k}_{i\bot},\xi_i$ to denote the variables describing the
motion of particle $i$ relative the system $1,2,...i-1,i+1,...N$. By
analogy with Eq. (\ref{3})
\begin{equation}
\xi_i=\frac{p_i^+}{P^+},\quad {\bf k}_{i\bot}={\bf p}_{i\bot}-\xi_i P^+
\label{22}
\end{equation}
Let $M_i$ be the free mass operator of the system
$1,2,...i-1,i+1,...N$ as a function of $\tilde{int}$. Then, by analogy
with Eq. (\ref{4}), we can introduce the vector
${\bf k}_i=({\bf k}_{i\bot},k_i^z)$ such that
\begin{equation}
\xi_i=\frac{\omega_i({\bf k}_i)+k_i^z}{M({\bf k}_i,M_i)},
\label{23}
\end{equation}
where $M({\bf k}_i,M_i)=\omega_i({\bf k}_i)+
(M_i^2+{\bf k}_i^2)^{1/2}$.

 By analogy with the two-particle case, we can introduce the internal
space $H_{int}$ as the space of functions
$\chi({\bf k}_{i\bot},\xi_i,\sigma_i,\tilde{int})$ such that
\begin{equation}
||\chi||^2\equiv \sum_{\sigma_i}\int\nolimits
|\chi({\bf k}_{i\bot},\xi_i,\sigma_i,\tilde{int})|^2
\frac{d^2{\bf k}_{i\bot}d\xi_i}{2(2\pi)^3\xi_i(1-\xi_i)}
d\rho(\tilde{int})\, < \, \infty
\label{24}
\end{equation}
Let the initial state of the system of $N$ particles be a bound state
with the total 4-momentum $P'$ and the internal wave function
$\chi'({\bf k}_{i\bot},\xi_i,\sigma_i,\tilde{int})$ the norm of which in
the space $H_{int}$ is equal to unity. Then the wave function of such a
system can be written in the form of Eq. (\ref{10}) (see, for example,
Ref. \cite{lev1}). After absorbing the
virtual photon with large momentum, particle $i$ becomes free, but the
rest of the system can consist of some number of free particles and of
some number of subsystems in the bound states. If
$\chi_i"(\tilde{int})$ is the internal wave function of the
system $1,2,...i-1,i+1,...N$ in the final state, then the wave function
of the system of $N$ particles in the final state can be written as
\begin{eqnarray}
&&|p_i",\sigma_i",P_i",\chi_i"\rangle =
2(2\pi)^3p_i^{"+}\delta^{(2)}({\bf p}_{i\bot}-{\bf p}_{i\bot}")
\delta(p_i^+-p_i^{"+})\delta_{\sigma_i\sigma_i"}\cdot\nonumber\\
&&2(2\pi)^3P_i^{"+}\delta^{(2)}({\bf P}_{i\bot}-{\bf P}_{i\bot}")
\delta(P_i^+-P_i^{"+})\chi_i"(\tilde{int})
\label{25}
\end{eqnarray}

 Instead of Eq. (\ref{13}) we should write the hadronic tensor in the
form
\begin{eqnarray}
&&W^{\mu\nu}=\frac{1}{4\pi}\sum_{\sigma_i"\tilde{int}}\int\nolimits
(2\pi)^4\delta^{(4)}(P'+q-p_i"-P_i")
\langle P',\chi'|J^{\mu}(0)|p_i",\sigma_i", \nonumber\\
&&P_i",\chi_i"\rangle \langle p_i",\sigma_i",P_i",\chi_i"|J^{\nu}(0)|
P',\chi'\rangle d\rho({\bf p}_{i\bot}",p_i^{"+})\cdot\nonumber\\
&&d\rho({\bf P}_{i\bot}",P_i^{"+})d\rho(\tilde{int})
\label{26}
\end{eqnarray}
where a sum is taken over all possible spin states of particle $i$ and
all possible internal states of the system $1,...i-1,i+1,...N$.

\begin{sloppypar}
 Let ${\bf k}_i"$ be the relative momentum of particle $i$ and the
system $1,...i-1,i+1,...N$, and $E_i({\bf d}_i)=(M_i^2+{\bf d}_i^2)^{1/2}$.
Then, as follows from Eqs. (\ref{10}),
(\ref{11}) and (\ref{25})
\begin{eqnarray}
&&\langle p_i",\sigma_i",P_i",\chi_i"(\tilde{int})|J_i^{\nu}(0)|
P',\chi'({\bf k}_i,\sigma_i,\tilde{int})\rangle=
\frac{r_i}{\xi_i}\sum_{\sigma_i'}\int\nolimits
\chi_i"(\tilde{int})^*\cdot\nonumber\\
&&[\bar{w}_i(p_i",\sigma_i")\gamma^{\mu}w_i(p_i',\sigma_i')]
\chi'({\bf d}_i,\sigma_i',\tilde{int})d\rho(\tilde{int})
\label{27}
\end{eqnarray}
where $p_i'=L[\beta({\bf P}_{\bot}'/M({\bf d}_i,M_i),
P^{'+}/M({\bf d}_i,M_i))]d_i$, and ${\bf d}_i$ is defined by the
condition, which can be written as Eq. (\ref{18}) if $M({\bf d}_i)$
is replaced by $M({\bf d}_i,M_i)$, $\tilde{\omega}_i$ is replaced by
$E_i$, and $M"=\omega({\bf k}")+E_i({\bf k}")$.
\end{sloppypar}

 Using the completeness of the states in the internal space of the
system $1,...i-1,i+1,...N$ and Eqs. (\ref{15}), (\ref{16}), (\ref{19}),
(\ref{26}) and (\ref{27}) we get (compare with Eq. (\ref{21}))
\begin{equation}
W^{\mu\nu}=\sum_{i=1}^{N} r_i^2 \int\nolimits \langle
\chi'({\bf k}_{i\bot},\xi_i=x,\tilde{int})|S_i^{\mu\nu}|
\chi'({\bf k}_{i\bot},\xi_i=x,\tilde{int})\rangle
\frac{d^2{\bf k}_{i\bot}d\rho(\tilde{int})}{4(2\pi)^3x(1-x)}
\label{28}
\end{equation}

 Let us introduce the notation
\begin{equation}
\rho_i(x)= \int\nolimits \langle
\chi'({\bf k}_{i\bot},\xi_i=x,\tilde{int})|\chi'({\bf k}_{i\bot},
\xi_i=x,\tilde{int})\rangle
\frac{d^2{\bf k}_{i\bot}d\rho(\tilde{int})}{2(2\pi)^3x(1-x)}
\label{29}
\end{equation}
Then, as follows from Eqs. (\ref{7}), (\ref{8}) and (\ref{24}),
$\rho_i(x)dx$ is
the probability of the event that particle $i$ in the bound state has
the value of $\xi_i$ in the interval $(x,x+dx)$.

 As follows from Eqs. (\ref{14}) and (\ref{28}), the structure functions
$F_1$ and $F_2$ depend only on $x$:
\begin{equation}
F_1(x)=\frac{1}{2}\sum_{i=1}^{N}r_i^2\rho_i(x),\quad F_2(x)=2xF_1(x)
\label{30}
\end{equation}
(the last equality is known as the Callan-Gross relation \cite{CallGr}).
These expressions for the structure functions were derived by many
authors in the framework of the parton model (see also Sec. \ref{S6}).
Equation (\ref{28}) also makes it possible to write the expression for the
polarized structure functions, but we shall not dwell on this question.

 One might think that the above results are natural since they fully
agree with the parton model. However the following question arises. Since
the ECO in the impulse approximation does not satisfy relativistic
invariance and current conservation (see the next section for more details),
the results for the structure functions depend on the reference frame
in which these functions are calculated. An argument in favor of choosing
the IMF is that in this reference frame the tensor $W^{\mu\nu}$ given by
Eq. (\ref{28}) satisfies the continuity equation
$q_{\mu}W^{\mu\nu}=q_{\nu}W^{\mu\nu}=0$.
Another well-known arguments are based on the approach proposed by
Weinberg \cite{Wein} and developed by several authors (see, for example,
Refs. \cite{BrLep,Namysl}). Let us note however that though quantum field
theory in the IMF seems natural and has some advantages, it also has some
serious difficulties which are not present in the usual formulation
\cite{GlWil}.

 In our opinion, a rather strange feature of the above results is as
follows. By looking through the derivation of these results one can
easily see that the initial state is treated in fact not as the bound
state but as the free state of noninteracting particles. Indeed, we have
never used the fact that the initial state is the eigenstate of the mass
operator ${\hat M}$ with the eigenvalue $M'$: ${\hat M}\chi'=M'\chi'$.
In the impulse approximation the relation between the quantities
${\bf d}_i$ and ${\bf k}_i"$ (see Eq. (\ref{18})) is derived from the
condition that the 4-vectors $(\tilde{\omega}({\bf d}_i),-{\bf d}_i)$
and $(\tilde{\omega}({\bf k}_i"),-{\bf k}_i")$ are connected by the
Lorentz boosts in the initial and final states. It is natural that
particle $i$ does not interact with the other particles in the final
state, but it is strange that we neglect the interaction in the initial
state and write the free mass $M({\bf d}_i)$ instead of the real mass
$M'$ which has the initial state.

 The effect of binding can be explicitly taken into account in models
where the ECO satisfies relativistic invariance and current conservation.
This problem is considered in the subsequent sections.

\section{Electromagnetic current operator for systems of interacting
particles}
\label{S4}

 In the following we use ${\hat  J}^{\mu}(x)$ to denote the ECO for a
system of interacting particles, while $J^{\mu}(x)$ is used to denote
the ECO in the impulse approximation.

Let ${\hat U}(a)=exp(\imath {\hat P}_{\mu}a^{\mu})$ be the
representation operator corresponding to the
displacement of the origin in spacetime translation of Minkowski
space by the 4-vector $a$. Here ${\hat P}=({\hat P}^0,{\hat {\bf
P}})$ is the operator of the 4-momentum, ${\hat P}^0={\hat E}$ is the
Hamiltonian, and ${\hat {\bf P}}$ is the operator of ordinary
momentum. Let also ${\hat U}(l)$ be the representation operator
corresponding to $l\in SL(2,C)$. Then ${\hat  J}^{\mu}(x)$ must be
the selfadjoint relativistic vector operator such that
\begin{equation}
{\hat U}(a)^{-1}{\hat  J}^{\mu}(x){\hat U}(a)=
{\hat J}^{\mu}(x-a)
\label{31}
\end{equation}
\begin{equation}
{\hat U}(l)^{-1}{\hat J}^{\mu}(x){\hat U}(l)=L(l)^{\mu}_{\nu}
{\hat J}^{\nu}(L(l)^{-1}x)
\label{32}
\end{equation}
where a sum over repeated indices $\mu,\nu=0,1,2,3$ is assumed.
As follows from Eq. (\ref{31}), the continuity equation
$\partial {\hat J}^{\mu}(x)/\partial x^{\mu}=0$ can be written in the
form
\begin{equation}
[{\hat  J}^{\mu}(x),{\hat P}_{\mu}]=0
\label{33}
\end{equation}
Since at least some of the operators ${\hat U}(a)$ and ${\hat U}(l)$
depend on interactions in the system under consideration, the immediate
consequence of Eqs. (\ref{31}-\ref{33}) is that ${\hat  J}^{\mu}(x)$
also depends on these interactions and thus ${\hat  J}^{\mu}(x)$ cannot
be written only as a sum of the constituent ECO's. This fact was first
pointed out by Siegert \cite{Sieg}.

 Let
\begin{equation}
\hat Q=\int\nolimits {\hat J}^{\mu}(x)d\sigma_{\mu}(x)
\label{34}
\end{equation}
be the system electric charge operator where $d\sigma_{\mu}(x)=
\lambda_{\mu}\delta(\lambda x-\tau)d^4x$ is the volume element of the
space-like hypersurface defined by the time-like vector $\lambda
\quad (\lambda^2=1)$ and the evolution parameter $\tau$. Then the
important physical condition is that the interactions do not
renormalize the electric charge, i.e. ${\hat Q}$ does not depend on
the choice of $\lambda$ and $\tau$ and has only one eigenvalue
equal to the sum of electric charges of constituents. It is well-known
that Eq. (\ref{33}) ensures that ${\hat Q}$ does not depend on
$\tau$ and $\lambda$ but this condition does not ensure that
${\hat Q}$ has the same value as for noninteracting particles.

 In addition, the operator ${\hat J}^{\mu}(x)$ should satisfy the
cluster separability condition. Briefly speaking, this condition
implies that if the interaction between any subsystems
$\alpha_1,...\alpha_n$ comprising the system under consideration is
turned off then ${\hat J}^{\mu}(x)$ must become a sum of the ECO's
${\hat J}_{\alpha_i}^{\mu}(x)$ for the subsystems.

 To explicitly construct the ECO satisfying the above properties it is
necessary to choose first the explicit realization of the
representation of the Poincare group for the
system under consideration. Dirac was the first who singled out three
forms of relativistic dynamics: instant, front and point ones
\cite{Dir}. As proved by Sokolov and Shatny \cite{SoSh}, these
forms are unitarily equivalent to each other. In Ref. \cite{lev} the
problem of constructing the ECO was first explicitly solved in the point
form, and then, using
the unitary operators constructed in Ref. \cite{SoSh}, the ECO was
constructed in the instant and front forms. For this reason, in the
present paper we use the solution in the point form. By definition,
the description in the point form implies that the operators
${\hat U}(l)$ are the same as for noninteracting particles, i.e.
${\hat U}(l)=U(l)$, and thus interaction terms can be present only in the
4-momentum operators ${\hat P}$ (i.e. in the general case
${\hat P}^{\mu}\neq P^{\mu}$ for all $\mu$).

\begin{sloppypar}
 In the point form it is convenient to use 4-velocities (instead of
4-momenta) as the external variables while the internal variables can
be chosen as above (see Eqs. (\ref{3}) and (\ref{6})). If $g_i=p_i/m_i$,
then it is easy to see that, by analogy with Eq. (\ref{7}),
\begin{equation}
d\rho({\bf g}_{1\bot},g_1^+)d\rho({\bf g}_{2\bot},g_2^+)=
d\rho({\bf G}_{\bot},G^+)d\rho(int)
\label{35}
\end{equation}
but now
\begin{equation}
d\rho(int)=\frac{M({\bf k})^2d^2{\bf k}_{\bot}d\xi}{2(2\pi)^3\xi(1-\xi)}
=\frac{M({\bf k})^3d^3{\bf k}}{2(2\pi)^3\omega_1({\bf k})
\omega_2({\bf k})}
\label{36}
\end{equation}
The internal two-particle Hilbert space $H_{int}$ can be formally defined
as the space of functions satisfying Eq. (\ref{8}), but with $d\rho(int)$
given by Eq. (\ref{36}).
\end{sloppypar}

 By analogy with the method proposed by Bakamdjian and Thomas in the
instant form \cite{BT}, it is possible to introduce the interaction into
the two-particle system as follows (see, for example, Refs.
\cite{sok,lev2}). First, we can express all the representation generators
of the Poincare group as functions of the operators $M$, $G$, and the
two-body spin operator ${\bf S}$. In this case $P=MG$, where $P$ is the free
two-body momentum operator, and the representation generators of the
Lorentz group are functions of only $G$ and ${\bf S}$. Then we replace $M$
by the two-body mass
operator ${\hat M}$ which acts only in $H_{int}$. If ${\hat M}$ commutes
with ${\bf S}$ then the commutation relations for the Poincare group
generators will not be broken. After this procedure the generators of the
Lorentz group remain the same as for the noninteracting particles, but
the 4-momentum operator ${\hat P}={\hat M}G$. In the general case the
two-body generators obtained in such a way should be subject to some
unitary transformation $A$, but we shall not discuss this question and
assume that $A=1$ (see Refs. \cite{lev3,lev} for more details).

 As follows from Eq. (\ref{31})
\begin{equation}
{\hat J}^{\mu}(x)=exp(\imath {\hat P}x){\hat J}^{\mu}(0)
exp(-\imath {\hat P}x)
\label{37}
\end{equation}
Therefore, if the operators ${\hat P}$ are known, it is sufficient to
construct only the operators ${\hat J}^{\mu}(0)$ with the correct
properties.

 Let $\varphi (G)$ be a function of $G$ with the range in $H_{int}$.
The action of ${\hat J}^{\mu}(0)$ in $H$ can be defined as
\begin{equation}
{\hat J}^{\mu}(0)\varphi (G)=2\int\nolimits {\hat M}^{3/2}{\hat J}^{\mu}
(G,G'){\hat M}^{3/2}\varphi (G')d\rho({\bf G}_{\bot}',G^{'+})
\label{38}
\end{equation}
where the kernel ${\hat J}^{\mu}(G,G')$ is an operator in $H_{int}$
for any fixed values of $G$ and $G'$.

 We use $\beta(G,G')$ to denote $\beta((G+G')/|G+G'|)\in$SL(2,C)
and $L(G,G')$ to denote $L[\beta(G,G')]$.
We also introduce the 4-vectors
\begin{equation}
f=L(G,G')^{-1}G, \qquad f'=L(G,G')^{-1}G'
\label{39}
\end{equation}
These 4-vectors are constructed as the c.m. frame 4-velocities of two
particles with unit masses and the 4-velocities $G$ and $G'$
(compare with Eq. (\ref{6})). Let us note that this is only a formal
construction since $G$ and $G'$ in Eq. (\ref{39}) have the sense of the
4-velocities of {\it one and the same} system in the final and
initial states. Nevertheless, as follows from Eq. (\ref{39}), the
4-vectors $f$ and $f'$ are such that
\begin{equation}
f^2=f'^2=1,\qquad {\bf f}+{\bf f}'=0,\qquad f^0=f'^0=(1+{\bf f}^2)^{1/2}
\label{40}
\end{equation}
Therefore the 4-vectors $f$ and $f'$ are fully determined by the spatial
part ${\bf f}$ of the 4-vector $f$.

  It can be shown (see Ref. \cite{lev}) that as follows from
Lorentz invariance
\begin{equation}
{\hat J}^{\mu}(G,G')=L(G,G')^{\mu}_{\nu}{\hat j}^{\nu}({\bf f})
\label{41}
\end{equation}
where we use ${\hat j}^{\nu}({\bf f})$ to denote ${\hat J}^{\nu}(f,f')$.
We see that the kernel of the operator ${\hat J}^{\mu}$ is fully
determined by an operator the action of which in $H_{int}$ depends only
on ${\bf f}$.

 The continuity equation (\ref{33}) in terms of ${\hat j}^{\nu}({\bf f})$
reads
\begin{equation}
f^0[{\hat M},{\hat j}^0({\bf f})]={\bf f}\{{\hat M},
{\hat {\bf j}}({\bf f})\}
\label{42}
\end{equation}
where we use curly brackets to denote the anticommutator. As shown in
Ref. \cite{lev}, the condition that the operator (\ref{34}) is the
same as for noninteracting particles will be satisfied if
${\hat j}^0(0)=j^0(0)$, i.e. the operator ${\hat j}^0(0)$ does
not depend on the interaction. Let us choose the coordinate axes in
such a way that ${\bf f}_{\bot}=0$. Then, as follows from Eq. (\ref{42}),
the continuity equation does not impose any constraint on the operator
${\hat {\bf j}}_{\bot}({\bf f})$. In addition, as shown in Ref.
\cite{lev}, Eq. (\ref{42}) makes it possible to find ${\hat j}^z({\bf f})$
if ${\hat j}^0({\bf f})$ is known. We conclude that one of the
possibilities to construct the ECO satisfying all the above properties
is to choose the operators ${\hat j}^0({\bf f})$ and
${\hat {\bf j}}_{\bot}({\bf f})$ in the same form as they have in the
case of noninteracting particles:
\begin{equation}
{\hat j}^{\nu}({\bf f})=j^{\nu}({\bf f})\quad {\mbox if} \quad \nu \neq z
\label{43}
\end{equation}
We shall assume that this condition is satisfied.

 As noted in the Introduction, there is also a problem, whether the ECO
satisfies the locality conditions, and in particular, the equal time
commutation relations. Generally speaking, we expect that these
conditions are not satisfied. However, as follows from Eqs. (\ref{37})
and (\ref{38}),
\begin{eqnarray}
&&{\hat J}^{\mu}(0,{\bf x}){\hat J}^{\nu}(0)\varphi (G)=
4exp(-\imath {\hat M}({\bf G}{\bf x}))
\int\nolimits\int\nolimits {\hat M}^{3/2}{\hat J}^{\mu}(G,G")
{\hat M}^3\cdot\nonumber\\
&&exp(\imath {\hat M}({\bf G}"{\bf x})){\hat J}^{\nu}(G",G')
\varphi (G')d\rho({\bf G}_{\bot}",G^{"+})d\rho({\bf G}_{\bot}',G^{'+})
\label{44}
\end{eqnarray}
If the kernel of the ECO is sufficiently smooth, then it follows from Eq.
(\ref{44}) that the strong limit of
${\hat J}^{\mu}(0,{\bf x}){\hat J}^{\nu}(0)$ is equal to zero if
$|{\bf x}|\rightarrow \infty$. Analogously it is easy to see that the
same is valid for the strong limit of
${\hat J}^{\nu}(0){\hat J}^{\mu}(0,{\bf x})$. As noted in the Introduction,
just these conditions should be necessarily satisfied.

 The explicit expression for the action of the free two-particle
operator $j^{\nu}({\bf f})$ in $H_{int}$ can be found from Eqs.
(\ref{38}-{40}), since in the case of noninteracting particles the ECO
is a sum of the ECO's for particles 1 and 2 (see Ref. \cite{lev} for more
details). The result is (compare with Eqs. (\ref{17}) and (\ref{18}))
\begin{eqnarray}
&&j^{\nu}({\bf f})\psi({\bf k},\sigma_1,\sigma_2)=
\sum_{i=1}^{2}\sum_{\sigma_i'} \frac{r_i}{2\omega({\bf d}_i)}
[\frac{M({\bf d}_i)}{M({\bf k}_i)}]^{3/2}\cdot\nonumber\\
&&[\bar{w}_i(h_i,\sigma_i)\gamma^{\nu}w_i(h_i',\sigma_i')]
\psi({\bf d}_i,\sigma_i',\tilde{\sigma}_i)
\label{45}
\end{eqnarray}
where
\begin{equation}
h_i'=L[\beta({\bf f}_{\bot}',f^{'+})](\omega_i({\bf d}_i),
{\bf d}_i), \quad h_i=L[\beta({\bf f}_{\bot},f^+)]
(\omega_i({\bf k}_i),{\bf k}_i),
\label{46}
\end{equation}
and instead of Eq. (\ref{18}), the vectors ${\bf d}_i$ are defined by
the conditions
\begin{equation}
L[\beta({\bf f}_{\bot}',f^{'+})](\tilde{\omega}_i({\bf d}_i),-{\bf d}_i)=
L[\beta({\bf f}_{\bot},f^+)](\tilde{\omega}_i({\bf k}_i),-{\bf k}_i)
\label{47}
\end{equation}

 As follows from Eqs. (\ref{5}) and (\ref{47})
\begin{equation}
f^{'+}[\tilde{\omega}_i({\bf d}_i)-d_i^z]=
f^+[\tilde{\omega}_i({\bf k}_i)-k_i^z],
\label{48}
\end{equation}
and, if ${\bf f}_{\bot}=0$,
\begin{equation}
{\bf d}_{\bot}={\bf k}_{\bot},\quad d_i^z=(1+2|{\bf f}|^2)k_i^z-
2f^0f^z\tilde{\omega}_i({\bf k}_i)
\label{49}
\end{equation}

 The problem of constructing the ECO for systems with $N>2$ particles
is much more complicated than in the case of two particles since
cluster separability imposes considerable restrictions on the choice
of the solution (see Ref. \cite{lev}). For this reason we shall consider
the case when (in the spirit of the parton model)
only the N-particle interaction is present
while there are no interactions in the subsystems of the system under
consideration. In this case the action of the operator $j^{\nu}({\bf f})$
in the N-particle internal space $H_{int}$ can be determined by analogy
with Eq.(\ref{45}):
\begin{eqnarray}
&&j^{\nu}({\bf f})\psi({\bf k}_i,\sigma_i,\tilde{int})=
\sum_{i=1}^{N}\sum_{\sigma_i'} \frac{r_i}{2\omega({\bf d}_i)}
[\frac{M({\bf d}_i,M_i)}{M({\bf k}_i,M_i)}]^{3/2}\cdot\nonumber\\
&&[\bar{w}_i(h_i,\sigma_i)\gamma^{\nu}w_i(h_i',
\sigma_i')]\psi({\bf d}_i,\sigma_i',\tilde{int})
\label{50}
\end{eqnarray}
where $h_i$ and $h_i'$ are defined by Eq.
(\ref{46}), $M_i$ is the free
mass of the system $1,...i-1,i+1,...N$, ${\bf d}_i$ is defined by
Eq. (\ref{47}) with $\tilde{\omega}_i({\bf d}_i)$ replaced by
$(M_i^2+{\bf d}_i^2)^{1/2}$, and for the index $i$ in the argument of the
wave function in the left-hand side we can take any integer from 1 to N.
By analogy with the two-particle case, we also assume that the N-particle
operator ${\hat j}^{\nu}({\bf f})$ satisfies Eq. (\ref{43}).

\section{Hadronic tensor for the system of N particles}
\label{S5}

 Let us first consider the problem of calculating the hadronic tensor
for the system of two particles. Let $g_i"$ $(i=1,2)$ be the 4-velocities
of the particles in the final state. Since the normalization of the
states should be the same as in Eq. (\ref{9}), the wave function of the
final state is
\begin{equation}
|g_1",\sigma_1",g_2",\sigma_2"\rangle =\prod_{i=1}^{2}\frac{2}{m_i}
(2\pi)^3g_i^{"+}\delta^{(2)}({\bf g}_{i\bot}-
{\bf g}_{i\bot}")\delta(g_i^+-g_i^{"+})\delta_{\sigma_i\sigma_i"}
\label{51}
\end{equation}

 Let $\psi'\in H_{int}$ be the internal wave function of the initial
bound state such that $||\psi'||^2=1$,
and $M'$ be the mass of this state. Then ${\hat M}\psi'=
M'\psi'$. If $P'$ is the 4-momentum of the initial state and $G'$ is
its 4-velocity then $P'=M'G'$, and, since the normalization should be
the same as for the wave function in Eq. (\ref{10}), we write
\begin{equation}
|G',\psi'\rangle =\frac{2}{M'}(2\pi)^3G^{'+}
\delta^{(2)}({\bf G}_{\bot}-{\bf G}_{\bot}')\delta(G^+-G^{'+})\psi'
\label{52}
\end{equation}

 Now the hadronic tensor should be written in the form
\begin{eqnarray}
&&W^{\mu\nu}=\frac{1}{4\pi}\sum_{\sigma_1"\sigma_2"}\int\nolimits
(2\pi)^4\delta^{(4)}(P'+q-p_1"-p_2")
\langle G',\psi'|{\hat J}^{\mu}(0)|g_1",\sigma_1", \nonumber\\
&&g_2",\sigma_2"\rangle
\langle g_1",\sigma_1",g_2",\sigma_2"|{\hat J}^{\nu}(0)|G',\psi'\rangle
d\rho({\bf p}_{1\bot}",p_1^{"+})d\rho({\bf p}_{2\bot}",p_2^{"+})
\label{53}
\end{eqnarray}

We again use Eqs. (\ref{6}), (\ref{15}) and (\ref{16}), and define the
4-velocity of the final state as $G"=P"/M"$. As follows from Eqs.
(\ref{35}) and (\ref{36}), the wave function given by Eq. (\ref{52})
can be rewritten in the form
\begin{eqnarray}
&&|g_1",\sigma_1",g_2",\sigma_2"\rangle = (2\pi)^3G^{"+}
\delta^{(2)}({\bf G}_{\bot}-{\bf G}_{\bot}")\delta(G^+-G^{"+})
\delta_{\sigma_1\sigma_1"}\delta_{\sigma_2\sigma_2"}\cdot\nonumber\\
&&\frac{2(2\pi)^3m_1m_2}{M({\bf k}")^3}\omega_1({\bf k}")
\omega_2({\bf k}")\delta^{(3)}({\bf k}-{\bf k}")
\label{54}
\end{eqnarray}
Therefore, as follows from Eqs. (\ref{35}), (\ref{36}), (\ref{38}),
(\ref{52}) and (\ref{54})
\begin{eqnarray}
&&\langle g_1",\sigma_1",g_2",\sigma_2"|{\hat J}^{\nu}(0)|G',
\psi'({\bf k},\sigma_1,\sigma_2)\rangle=\frac{2}{M'm_1m_2}
(M"M')^{3/2}\cdot\nonumber\\
&&{\hat J}^{\nu}(G",G')\psi'({\bf k}",\sigma_1",\sigma_2")
\label{55}
\end{eqnarray}

 Since the ECO satisfies relativistic invariance and current conservation,
the results for the structure functions do not depend on the reference
frame in which $W^{\mu\nu}$ is calculated, and in any frame
$q_{\mu}W^{\mu\nu}=q_{\nu}W^{\mu\nu}=0$. As follows from Eqs.
(\ref{39}-{\ref{41}), it is convenient to choose the reference frame in
which ${\bf G}"+{\bf G}'=0$, since in this case the Lorentz transformation
$L(G",G')$ is the identity operator, and thus
${\hat J}^{\mu}(G,G')={\hat j}^{\mu}({\bf f})$ with ${\bf f}={\bf G}"$.
Let us note that the condition ${\bf G}"+{\bf G}'=0$ is not the same as
the condition ${\bf P}"+{\bf P}'=0$ defining the Breit frame, since the
masses of the initial and final states are different.

 We again suppose that ${\bf P}_{\bot}'={\bf q}_{\bot}=0$ and $P^{'z}>0$.
Therefore ${\bf G}_{\bot}'={\bf G}_{\bot}"=0$, and $G^{'z}>0$. If
${\bf G}"+{\bf G}'=0$ then $f^z=G^{"z}<0$, and, since
${\bf G}"=(M'{\bf G}'+{\bf q})/M"$, we find that
\begin{equation}
q^0=(M"-M')G^{'0},\quad {\bf q}=-(M"+M'){\bf G}'
\label{56}
\end{equation}
As follows from these expressions
\begin{equation}
{\bf G}^{'2}=|{\bf f}|^2=\frac{(M"-M')^2-q^2}{4M"M'}
\label{57}
\end{equation}
In the Bjorken limit $M"\gg M'$, and, since $M^{"2}=(P'+q)^2$, we get
from Eqs. (\ref{16}) and (\ref{57})
\begin{equation}
|{\bf f}|^2=\frac{|q^2|^{1/2}}{4M'[x(1-x)]^{1/2}}
\label{58}
\end{equation}
Therefore $|{\bf f}|\gg 1$ in the Bjorken limit.

 Using Eqs. (\ref{43}), (\ref{45}), (\ref{46}) and (\ref{55}), we obtain
that if $\nu\neq z$, then in the reference frame under consideration
\begin{eqnarray}
&&\langle g_1",\sigma_1",g_2",\sigma_2"|{\hat J}^{\nu}(0)|G',
\psi'({\bf k},\sigma_1,\sigma_2)\rangle
\sum_{i=1}^{2}\sum_{\sigma_i'} \frac{r_iM({\bf d}_i)^{3/2}(M')^{1/2}}
{2\omega({\bf d}_i)m_1m_2}\cdot\nonumber\\
&&[\bar{w}_i(h_i",\sigma_i")\gamma^{\nu}w_i(h_i',
\sigma_i')]\psi({\bf d}_i,\sigma_i',\tilde{\sigma}_i")
\label{59}
\end{eqnarray}
where $h_i'$, $h_i"=h_i$ and ${\bf d}_i$
are defined by Eqs. (\ref{46}) and (\ref{47}) with ${\bf f}_{\bot}'=
{\bf f}_{\bot}=0$, ${\bf k}_{i\bot}={\bf k}_{i\bot}"$.

 As follows from Eq. (\ref{49}), in the reference frame under
consideration
\begin{equation}
d_i^z=(1+2|{\bf f}|^2)[k_i^{"z}-\tilde{\omega}_i({\bf k}_i")]-
\frac{\tilde{\omega}_i({\bf k}_i")}{4|f^z|^2}
\label{60}
\end{equation}
Since $\tilde{\omega}_i({\bf k}_i")/4|f^z|^2=M'(1-x)/2$ (see Eqs.
(\ref{16}) and (\ref{58})), we conclude that the condition
$|{\bf d}|\leq m_0$ (see Sec. \ref{S2}) can be again satisfied only if
$1-|cos\theta|\leq m_0^2/|q^2|$ for $i=1$ and
$1+cos\theta\leq m_0^2/|q^2|)$ for $i=2$. Therefore, the presence of
the interaction in the ECO does not change the conclusion that (at least
if $\mu,\nu \neq z$) the constituents absorb the virtual photon
incoherently.

 Let us now consider Eq. (\ref{48}). As follows from Eq. (\ref{4}),
$\tilde{\omega}_i({\bf d}_i)-d_i^z=M({\bf d}_i)(1-\xi_i)$. Taking into
account the fact that the free mass operator can also be written as a
function of ${\bf d}_{i\bot}$ and $\xi_i$, and using Eq. (\ref{49}), we can
write $M({\bf d}_i)=M({\bf k}_{i\bot}",\xi_i)$. We also take into account
that in the reference frame under consideration
\begin{equation}
f^{'+}=\sqrt{2}|f^z|,\quad f^+=\frac{1}{2\sqrt{2}|f^z|}
\label{61}
\end{equation}
Therefore, using Eqs. (\ref{16}) and (\ref{58}) we get the final result
\begin{equation}
M({\bf k}_{i\bot}",\xi_i)(1-\xi_i)=M'(1-x)
\label{62}
\end{equation}
where, as can be shown from Eq. (\ref{4}), the explicit expression for
$M({\bf k}_{i\bot}",\xi_i)$ is
\begin{equation}
M({\bf k}_{i\bot}",\xi_i)=[\frac{m_1^2}{\xi_1}+\frac{m_2^2}{\xi_2}+
\frac{{\bf k}_{\bot}^{"2}}{\xi_1\xi_2}]^{1/2}
\label{63}
\end{equation}

 We see that the equality $\xi_i=x$ takes place only if one neglects the
difference between the free mass and the mass of the bound state. This
equality was obtained from Eq. (\ref{18}) while the relation (\ref{62})
was obtained from Eq. (\ref{47}) at ${\bf k}_i"={\bf k}_i$. Since in the
reference frame under consideration $P'=M'f',\, P"=M"f$, we can rewrite
Eq. (\ref{47}) in the form
\begin{equation}
L[\beta(\frac{{\bf P}_{\bot}'}{M'},\frac{P^{'+}}{M'})]
(\tilde{\omega}_i({\bf d}_i),-{\bf d}_i)=
L[\beta(\frac{{\bf P}_{\bot}"}{M"},\frac{P^{"+}}{M"})]
(\tilde{\omega}_i({\bf k}_i"),-{\bf k}_i")
\label{64}
\end{equation}
Therefore, instead of the free mass $M({\bf d}_i)$ in Eq. (\ref{18}),
the mass of the initial bound state $M'$ enters into Eq. (\ref{64}).
This explains the result given by Eq. (\ref{62}).

 An analogous effect (called $x$-rescaling) was observed by the authors
investigating the original EMC effect \cite{EMC}, and in Sec. \ref{S6}
we discuss the difference between our results and those in Refs.
\cite{Ak,FS,Ciofi}.

 Equation (\ref{64}) has the clear physical meaning. Indeed,
the Lorentz boost in the left-hand side is the real physical boost
since $P'$ and $M'$ are the real 4-momentum and the mass which has the
initial state. Let the virtual photon be absorbed by particle 1. Then
the left-hand side of Eq. (\ref{64}) $h_2'$ has the meaning of
the momentum of particle 2 in the initial state since the 4-momentum of
this particle in the c.m. frame of the initial state is
$(\omega_2({\bf d}_1),-{\bf d}_1)$. Analogously, the right-hand side
of Eq. (\ref{64}) $h_2"$ has the meaning of the momentum of
particle 2 in the final state. Thus Eq. (\ref{64}) tells that the
momentum of particle 2 does not change. As follows from Eqs. (\ref{6})
and (\ref{47}), $h_1"=p_1"$ and $h_2"=p_2"$ since in
the final state the two-particle system is free. At the same time, the
quantities $h_1'$ and $h_2'$ are not equal to the
free momenta $p_1'$ and $p_2'$ in the initial state since
$h_1'$ and $h_2'$ are defined by the Lorentz boost
depending on the physical mass $M'$ while $p_1'$ and $p_2'$ are defined
by the Lorentz boost depending on the free mass $M({\bf d}_1)$.
Meanwhile, Eq. (\ref{18}) just tells that $p_2'=p_2"$ as it should be
from the definition of the impulse approximation. We conclude that
since in the presence of the interaction the quantity $h_2'$
can be interpreted as the physical 4-momentum of particle 2 in the
initial state, while $p_2'$ no longer can be interpreted in such a way,
Eq. (\ref{64}) is reasonable while Eq. (\ref{18}) is not.

 Since we wish to compare the results with those obtained in the
impulse approximation, we note that, as follows from Eqs. (\ref{7})
and (\ref{36})
\begin{equation}
\psi'({\bf d},\sigma_1,\sigma_2)=\frac{m_1m_2}{M({\bf d})}
\chi'({\bf d},\sigma_1,\sigma_2)
\label{65}
\end{equation}
Then a simple calculation using Eqs. (\ref{4}), (\ref{15}), (\ref{16}),
(\ref{46}), (\ref{53}), (\ref{58}-{60}) shows that in the reference frame
under consideration the components of the tensor $W^{\mu\nu}$ with $\mu,\nu
\neq z$ are given by
\begin{equation}
W^{\mu\nu}=\sum_{i=1}^{2} r_i^2 \int\nolimits \langle
\chi'({\bf k}_{i\bot},\xi_i)|S_i^{\mu\nu}|
\chi'({\bf k}_{i\bot},\xi_i)\rangle [1+\frac{k_i^z}{\omega_i({\bf k})}]^2
\frac{d^2{\bf k}_{i\bot}}{4(2\pi)^3\xi_i(1-x)}
\label{66}
\end{equation}
where, as follows from Eqs. (\ref{62}) and (\ref{63}), $\xi_i$ is a
function of ${\bf k}_{i\bot}$ and $x$ which should be defined from the
condition
\begin{equation}
[\frac{m_i^2}{\xi_i}+\frac{{\tilde m}_i^2}{1-\xi_i}+
\frac{{\bf k}_{\bot}^2}{\xi_i(1-\xi_i)}]^{1/2}(1-\xi_i)=M'(1-x)
\label{67}
\end{equation}

 By analogy with the calculations in Sec. \ref{S3}, we can easily
generalize the above calculations to the case of N particles in the
model considered in Sec. \ref{4} where ${\hat j}^({\bf f})$ is
given by Eqs. (\ref{50}) and (\ref{43}). In the reference frame where
${\bf P}_{\bot}'={\bf P}_{\bot}"=0$, $P^{'z}>0$ and
${\bf G}'+{\bf G}"=0$,
\begin{eqnarray}
W^{\mu\nu}&=&\sum_{i=1}^{N} r_i^2 \int\nolimits \langle
\chi'({\bf k}_{i\bot},\xi_i,\tilde{int})|S_i^{\mu\nu}|
\chi'({\bf k}_{i\bot},\xi_i,\tilde{int})\rangle \cdot\nonumber\\
&&[1+\frac{k_i^z}{\omega_i({\bf k})}]^2
\frac{d^2{\bf k}_{i\bot}d\rho(\tilde{int})}{4(2\pi)^3\xi_i(1-x)}
\label{68}
\end{eqnarray}
where $\xi_i$ is a function of ${\bf k}_{i\bot},M_i,x$ defined
by the equations
\begin{eqnarray}
&&M({\bf k}_{i\bot},\xi_i,M_i)(1-\xi_i)=M'(1-x),\nonumber\\
&&M({\bf k}_{i\bot},\xi_i,M_i)=[\frac{m_i^2}{\xi_i}+\frac{M_i^2}{1-\xi_i}+
\frac{{\bf k}_{\bot}^2}{\xi_i(1-\xi_i)}]^{1/2}
\label{69}
\end{eqnarray}
and $k_i^z$ is a function of ${\bf k}_{i\bot},M_i,x$ defined by
Eq. (\ref{23}).

 It is easy to see that the explicit expression for $\xi_i$ is
\begin{eqnarray}
&&\xi_i=\frac{m_i^2+{\bf k}_{i\bot}^2}{m_i^2+{\bf k}_{i\bot}^2+
M^{'2}(1-x)^2}\quad {\mbox if}\quad m_i=M_i,\nonumber\\
&&\xi_i=\frac{1}{2}\{1-\alpha_i-\beta_i+[(1-\alpha_i-\beta_i)^2+
4\alpha_i]^{1/2}\}\quad {\mbox if} \quad M_i>m_i, \nonumber\\
&&\xi_i=\frac{1}{2}\{1-\alpha_i-\beta_i-[(1-\alpha_i-\beta_i)^2+
4\alpha_i]^{1/2}\}\quad {\mbox if} \quad M_i<m_i
\label{70}
\end{eqnarray}
where
\begin{equation}
\alpha_i=\frac{m_i^2+{\bf k}_{i\bot}^2}{M_i^2-m_i^2},\quad
\beta_i=\frac{M^{'2}(1-x)^2}{M_i^2-m_i^2}
\label{71}
\end{equation}
Since $x\in [0,1]$, it follows from Eq. (\ref{70}) that
$\xi_i \in [\xi_i^{min},1]$ where $\xi_i^{min}=\xi_i^{min}({\bf
k}_{i\bot},M_i)$ is a function of ${\bf k}_{i\bot},M_i$ which can be
defined from Eq. (\ref{70}) at $x=0$. It is easy to see that
$0<\xi_i^{min}<1$.

 As follows from Eqs. (\ref{14}) and (\ref{68}), the scaling and the
Callan-Gross relation \cite{CallGr} also take place if the interaction
in the ECO is taken into account since
\begin{eqnarray}
&&F_1(x)=\sum_{i=1}^{N} r_i^2 \sum_{\sigma_i} \int\nolimits
|\chi'({\bf k}_{i\bot},\xi_i,\sigma_i,\tilde{int})|^2
[1+\frac{k_i^z}{\omega_i({\bf k}_i)}]^2\frac{d^2{\bf k}_{i\bot}
d\rho(\tilde{int})}{4(2\pi)^3\xi_i(1-x)},\nonumber\\
&&F_2(x)=2xF_1(x)
\label{72}
\end{eqnarray}

\section{Discussion}
\label{S6}

 Though there exist a vast literature devoted to the parton model, only
a few authors investigated the problem, what explicit physical conditions
should be satisfied for the validity of this model (see, for example,
Refs. \cite{Close,ILH,Web,Drell}). The results of Secs. \ref{S2} and
\ref{S3} show that the impulse approximation is the sufficient condition
for ensuring the validity of the parton model in the Bjorken limit
(in agreement with the above references). It is obvious from
Eqs. (\ref{3}) and (\ref{22}) that the quantity $\xi_i$ is indeed the
fraction of the total momentum in the IMF which has the particle
interacting with the virtual photon, and the results show that
indeed $\xi_i=x$ in the Bjorken limit.

 It is easy to show that if the point-like particle $i$ with the spin
1/2 and the initial 4-momentum $\xi_iP'$ absorbs the virtual photon with
the 4-momentum $q$, then $F_{2i}(x/\xi_i)=\delta(x/\xi_i-1)$, and
therefore, as follows from Eq. (\ref{30}),
\begin{equation}
F_2(x)=\sum_{i=1}^{N}r_i^2\int_{0}^{1} F_{2i}(\frac{x}{\xi_i})
\rho_i(\frac{x}{\xi_i})d\xi_i
\label{73}
\end{equation}
in full agreement with the interpretation of the function
$\rho_i(\xi_i)$ (see Sec. \ref{S3}) and with the expression used by many
authors.

 The general expression for the hadronic tensor is
\begin{equation}
W^{\mu\nu}=\frac{1}{4\pi}\sum_{n}(2\pi)^4\delta^{(4)}(P'+q-P_n)
\langle P',\chi'|{\hat J}^{\mu}(0)|n\rangle
\langle n|{\hat J}^{\nu}(0)|P',\chi'\rangle
\label{74}
\end{equation}
where a sum is taken over all possible intermediate states $|n\rangle$,
and $P_n$ is the 4-momentum of the state $|n\rangle$. It is well-known
that using Eq. (\ref{37}) and the completeness of the states $|n\rangle$,
it is easy to transform Eq. (\ref{74}) to the form
\begin{equation}
W^{\mu\nu}=\frac{1}{4\pi}\int\nolimits e^{\imath qx} \langle P',\chi'|
{\hat J}^{\mu}(x){\hat J}^{\nu}(0)|P',\chi'\rangle d^4x
\label{75}
\end{equation}
The product of the ECO's in this expression can also be replaced by the
commutator since the second term in the commutator does not contribute
to the integral.

 The usual argument in favor of the impulse approximation is that since
at large $q$ only the region of small $x$ contributes to the integral
(see, for example, Ref. \cite{Ioffe}), the asymptotic freedom guarantees
that ${\hat J}^{\mu}(x)$ can be replaced by the free ECO $J^{\mu}(x)$
with a good accuracy. Then using again the completeness of the states
$|n\rangle$, we obtain that the hadronic tensor will be replaced by
the following expression
\begin{equation}
W^{\mu\nu}\rightarrow \frac{1}{4\pi}\sum_{n}(2\pi)^4
\delta^{(4)}(P^{'(0)}+q-P_n^{(0)})\langle P',\chi'|J^{\mu}(0)|n\rangle
\langle n|J^{\nu}(0)|P',\chi'\rangle
\label{76}
\end{equation}
Here $P^{'(0)}$ and $P_n^{(0)}$ are not the real 4-momenta in the initial
state and in the state $|n\rangle$, but the total 4-momenta of the
{\it free} constituents comprising these states.

 It is clear from the considerations in Secs. \ref{S2} and \ref{S5}, that
the replacement $P_n\rightarrow P_n^{(0)}$ is reasonable at least in some
cases. The argument in favor of the replacement $P'\rightarrow P^{'(0)}$
is also well-known: the quantum theory on the light cone is such that
the $\bot$ and $+$ components of the vectors $P'$ and $P^{'(0)}$ are equal
to each other and only the "minus" components differ, but these components
are equal to zero in the IMF. Therefore, as far as the $x$ dependence of
${\hat J}^{\mu}(x)$ is concerned, the effect of binding is indeed
negligible. At the same time we do not see the reason why
${\hat J}^{\mu}(0)$ can be replaced by $J^{\mu}(0)$, and this is just
the impulse approximation. If we expand ${\hat J}^{\mu}(0)$ in powers
of $\alpha_s$, the same should be done with the initial state, but the
perturbation theory cannot be used in this case.

 In our opinion, the crucial point in understanding the situation is that
not only the 4-momentum of the initial state, {\it but also the mass of
this state} enter into the calculations. Therefore we cannot confine
ourselves to the consideration of only the quark absorbing the virtual
photon, and the large distances necessarily come into play. The impulse
approximation unambiguously leads to the prescription that for the mass
of the initial state we should take not the physical mass $M'$, but the
mass of the system of free constituents, i.e. the nonphysical quantity.
In contrast with the case of the "minus" components of the vectors $P'$
and $P^{'(0)}$ in the IMF, we cannot make the difference between the
above masses negligible.

 In our calculations in Sec. \ref{S5} we used the ECO which satisfies
relativistic invariance and current conservation, but, as noted above,
these conditions are not sufficient for choosing a unique solution.
Nevertheless the ECO under consideration unambiguously leads to the
prescription that for the mass of the initial state we should take its
physical value $M'$. Therefore, we believe, that though our result
(\ref{68}) for the hadronic tensor is model-dependent, the relation
between $\xi_i$ and $x$ given by Eqs. (\ref{67}), (\ref{69}-\ref{71})
does not depend on the choice of the solution for the ECO
if the quarks absorb the virtual photon
incoherently. Indeed, as explained in Sec. \ref{S5}, these expressions
are only the consequence of relativistic kinematics. For this reason
we expect that in the general case the relations (\ref{69}-{71}) will
be valid if $M_i$ is replaced by the mass operator ${\hat M}_i$ of the
subsystem $1,...i-1,i+1,...N$. However, to prove this statement it is
necessary to construct the ECO in the case when the interactions in the
subsystems of the system under consideration are present.

 In the literature the impulse approximation is often associated with
the Feynman diagrams in which the virtual photon interacts with only
one constituent, while the other constituents are spectators.
Let us note however that each Feynman diagram can be unambiguously
calculated only if the underlying dynamics is known (for both the
interaction between the constituents and the ECO satisfying Eqs.
(\ref{31}-{33})). Meanwhile usually this is not the case, and the
Feynman diagrams are calculated using some prescriptions. Our solution
for the ECO unambiguously leads to Eq. (\ref{47}) which shows that for
the particle which does not interact with the virtual photon
$h_{initial}=p_{final}$ (see Sec. \ref{S5}). This looks like
the impulse approximation. However, as explained in Sec. \ref{S5}, the
quantity $h_{initial}$ is not equal to the free 4-momentum
$p_{initial}$. The difference between these quantities cannot be
described in the perturbation theory. So it is not clear what is the
interpretation of our result on the language of Feynman diagrams.

As follows from Eq. (\ref{69}), the relation $\xi_i=x$ takes place
only in the nonrelativistic approximation. Therefore we should expect
that in the real nucleon $\xi_i$ considerably differs from $x$. Thus,
in contrast with Eq. (\ref{30}), for determining the structure functions
it is necessary to know not only the functions $\rho_i(\xi_i)$, but
also the dependence of the internal wave function on the transverse
momenta. As the result, the DIS data do not make it possible to
determine the $\xi_i$ distribution of quarks in the nucleon if there
are no additional experimental information.  We can also
expect that the sum rules which are based only on the parton model are
not reliable.

 Let us consider, for example, the Gottfried sum rule \cite{Got},
according to which the quantity
\begin{equation}
S_G=\int\nolimits [F_{2p}(x)-F_{2n}(x)]\frac{dx}{x}
\label{77}
\end{equation}
is equal to 1/3. Here $F_{2p}(x)$ and $F_{2n}(x)$ are the structure
functions for the proton and neutron respectively. This sum rule
easily follows from Eqs. (\ref{24}), (\ref{29}), (\ref{30}) and the
condition $||\chi'||=1$ if we assume that the neutron wave function
can be obtained from the proton one if one of the $u$ quarks in the
proton is replaced by the $d$ quark (though some authors argue that
this is not the case).
 We suppose that particle 1 in the proton is the $u$ quark, particle
1 in the neutron is the $d$ quark and all other particles are the same.
Then, as follows from Eq. (\ref{72})
\begin{equation}
S_G=\frac{1}{3} \sum_{\sigma_i} \int\nolimits
|\chi'({\bf k}_{1\bot},\xi_1,\sigma_1,\tilde{int})|^2
[1+\frac{k_1^z}{\omega_1({\bf k}_1)}]^2\frac{d^2{\bf k}_{1\bot}
d\rho(\tilde{int})dx}{2(2\pi)^3\xi_1(1-x)}
\label{78}
\end{equation}
where $\xi_1$ is a function of ${\bf k}_{1\bot},M_1,x$ defined by Eq.
(\ref{70}). Now using Eqs. (\ref{23}) and (\ref{69}-{71}) we change the
integration variable from $x$ to $\xi_1$.  Then it is easy to show that
\begin{eqnarray}
S_G&=&\frac{1}{3} \sum_{\sigma_i} \int\nolimits
\frac{d^2{\bf k}_{1\bot}d\rho(\tilde{int})}{(2\pi)^3}
\int_{\xi_1^{min}}^{1} \frac{d\xi_1}{2\xi_1(1-\xi_1)}\cdot\nonumber\\
&&|\chi'({\bf k}_{1\bot},\xi_1,\sigma_1,\tilde{int})|^2
[1+\frac{k_1^z}{\omega_1({\bf k}_1)}]
\label{79}
\end{eqnarray}
It is not clear what is the effect of the last multiplier in the
integrand since both $k_1^z>0$ and $k_1^z<0$ are possible. However in
the general case $\xi_1^{min}$ can considerably differ from zero.
Therefore, comparing Eq. (\ref{79}) with the normalization integral
(\ref{24}), it is natural to expect that $S_G<1/3$. Recently the
quantity $S_G$ was calculated in Ref. \cite{NMC1} using the data of Ref.
\cite{NMC}, and the result was $S_G=0.235\pm 0.026$.

Analogously, the DIS data only do not make it possible to determine the
contributions of the $u$, $d$ and $s$ quarks to the nucleon spin, and the
well-known problem of the "spin crisis" does not arise (the present
status of this problem is described, for example, in Refs.
\cite{Ellis,Ioffe1}). Indeed, these contributions (usually denoted as
$\Delta q=(\Delta u,\Delta d,\Delta s)$) are given by some integrals
over $\xi_i \in [0,1]$. Meanwhile, the DIS data make it possible to
calculate some integrals over $x\in [0,1]$. Since the integrals over
$x$ can be transformed to the integrals over
$\xi_i \in [\xi_i^{min},1]$, we see that the DIS data do not make it
possible to determine the contributions of $\xi_i \in [0,\xi_i^{min}]$.
Thus it is natural to expect that the parton model underestimates the
quantities $\Delta q$.

At the same time, the DIS experiments make it possible to
check the well-known results which are not based on the parton model
(for example, the Bjorken sum rules \cite{Bjor1}).

 In conclusion we compare our results with those obtained by several
authors investigating the original EMC effect \cite{EMC}. This is
possible in the formal case when the nucleons are point-like.

 The first calculations of the EMC effect were carried out in Refs.
\cite{Ak} and others. However, as shown in Ref. \cite{FS}, the above
works did not take into account the "flux factor" (see also Refs.
\cite{Ciofi} and many others). The flux factor used in these
references is equal to $z_i=A(\omega_i({\bf k}_i)-k_i^z)/M_A$ where
$A$ is the mass number of the nucleus under consideration and $M_A$ is
its mass (equal to $M'$ in our notations).

 If we work in the impulse approximation then Eq. (\ref{73}) is the
convolution formula for $F_2(x)$. Since $x/\xi_i=Ax/A\xi_i$, we
conclude that the result for the flux parameter in the
impulse approximation is $A\xi_i$ where $\xi_i$ is given by Eq.
(\ref{23}). Note however, that since the ECO in the impulse approximation
is not relativistically invariant, the result depends on the reference
frame and on the form of dynamics (see Ref. \cite{Coester} for more
details). We derived our result in the IMF while the EMC effect is usually
considered in the reference frame in which the initial nucleus is at rest.

 Let us now consider the case when the interaction in the ECO is taken
into account. Then if $N=2$, our result which follows from Eq.
(\ref{63}) is $\tilde{\omega}_i({\bf k}_i)-k_i^z=M'(1-x)$ while in the
approach of Refs. \cite{FS,Ciofi} the relation
$\omega_i({\bf k}_i)-k_i^z=M'x$ takes place. If $N$ is arbitrary, then
our result (\ref{69}) can be written as
$(M_i^2+{\bf k}_i^2)^{1/2}-k_i^z=M'(1-x)$ while the result of the above
references is again $\omega_i({\bf k}_i)-k_i^z=M'x$.

 Above we have argued that Eqs. (\ref{62}) and (\ref{69}) are in fact
kinematical. In addition, it is easy to see that the expression for
$F_2(x)$ in Eq. (\ref{72}) cannot be written as the one-dimensional
convolution formula. This can be expected in any model in which the
interaction in the ECO is taken into account (see also Refs.
\cite{Ciofi,Mel}).

 These considerations show that the interpretation of the original EMC
effect \cite{EMC} has to be revisited.  We suppose to consider this
problem elsewhere.

\vskip 1em
\begin{center} {\bf Acknowledgments} \end{center}
\vskip 0.5em
\begin{sloppypar}
 The author is grateful to S.B.Gerasimov, S.Gevorkyan, I.L.Grach,
A.V.Efremov,  E.A.Kuraev, G.I.Lykasov, L.P.Kaptari, A.Makhlin,
S.V.Mihailov, I.M.Narodetskii, M.G.Sapozhnikov, N.B.Skachkov,
Y.N.Uzikov  and  H.J.Weber for valuable discussions. This work was
supported by grant No. 93-02-3754 from the Russian Foundation for
Fundamental Research.
\end{sloppypar}

\end{document}